\pdfoutput=1
\documentclass[fontsize=9pt, DIV=calc, a4paper, twocolumn, headings=standardclasses]{scrartcl}

\usepackage[utf8]{inputenc}
\usepackage[english]{babel}
\usepackage{lmodern}
\usepackage[activate={true, nocompatibility}, final, tracking=true, kerning=true, factor=1100, stretch=10, shrink=10]{microtype}
\usepackage{booktabs, graphicx, amsmath, amssymb, physics, numprint, pdfpages}
\usepackage[format=plain, labelfont=bf]{caption}
\usepackage[auth-lg, affil-it]{authblk}
\usepackage[sort&compress, numbers]{natbib}
\usepackage[hidelinks]{hyperref}

\npthousandsep{,}
\DeclareMathOperator{\SD}{SD}
\DeclareMathOperator{\std}{std}
\DeclareMathOperator{\mean}{mean}
\newcommand{\ccom}[1]{\textbf{#1}}
\newcommand{\orig}{\textit{Zagorski et al.:} }
\newcommand{\reply}{\textit{Reply:} }

\title{Reply to: Assessing the precision of morphogen gradients\\in neural tube development}

\author[$1,2$]{Roman Vetter}
\author[$1,2$]{Dagmar Iber}

\affil[$1$]{Department of Biosystems Science and Engineering, ETH Z\"{u}rich, Schanzenstrasse 44, 4056 Basel, Switzerland}
\affil[$2$]{Swiss Institute of Bioinformatics, Schanzenstrasse 44, 4056 Basel, Switzerland}

\date{\today}

\begin{document}

\twocolumn[
\begin{@twocolumnfalse}
\maketitle
\begin{abstract}
\hrule
\vskip0.5\baselineskip
\textit{This document supersedes the \href{https://doi.org/10.1038/s41467-024-45149-7}{version published on 1 February 2024 in Nat.\ Commun.\ 15, 930}. The Matters Arising piece by Zagorski et al., that we were invited to reply to, differed substantially from the \href{https://doi.org/10.1038/s41467-024-45148-8}{version published in Nat.\ Commun.\ 15, 929}. The editorial process did not allow us to address these changes in our reply published in Nat.\ Commun.}
\vskip0.5\baselineskip
\hrule
\end{abstract}
\vskip\baselineskip
\end{@twocolumnfalse}
]{}

\noindent\textbf{In a recent article \cite{Vetter:2022}, we demonstrated that single morphogen gradients in the developing mouse neural tube (NT) can carry sufficient positional accuracy to explain the patterning precision of progenitor domain boundaries. Zagorski et al.\ had previously concluded otherwise \cite{Zagorski:2017}, based on methodological inconsistencies that we have revealed. The authors now comment on our work with a Matters Arising letter. We rebut their criticism point by point in the Supplement, and summarize the main aspects here.}

\vskip\baselineskip

The authors criticize our comparison of the different analysis methods (FitEPM, NumEPM, DEEM) using a set of exponential functions. It is a fundamental part of quantitative science to validate all methods used to process data—be it part of experimental, theoretical or numerical work—against known results. If a method fails to provide the correct results for known problems, it is unreasonable to apply it to similar problems, for which the answer is not known. The authors' indirect approximation (FitEPM) leads to a vast overestimation of the positional error in case of exponential gradients, and is accurate only near the morphogen source \cite{Vetter:2022}. The same limitations apply to noisy gradient data, and the challenges of background subtraction and smoothening apply to all three methods.

The authors insist that further away from the source the observed gradients are flat, such that our arguments would not apply. First, their approximation (FitEPM) will yield the wrong positional error whenever the mean deviates substantially from an exponential \cite{Vetter:2022} (i.e., also in flat parts of the gradient), as it makes use of the exponential shape explicitly. Second, the flat part of the gradients is not reliable biological data. We speculated in our paper that the switch to a flat gradient shape may be due to insufficient imaging depth, but the employed imaging depth remained unknown to us \cite{Vetter:2022}. We have since received confirmation that Zagorski et al.\ indeed employed 8-bit imaging (A.~Kicheva, personal communication, Supplementary Information). 8-bit imaging only permits the detection of a $2^8=256$-fold intensity change. As such, it is technically impossible to detect an exponential gradient beyond $5.5$ times its decay length ($\approx110$\,\textmu{}m) from the source, which coincides with the point where Zagorski et al.\ find the transition from an exponential to a flat gradient shape. The physical limits of their imaging setup and the mathematical limitation of their approximation of the positional error make it impossible to evaluate the positional error of the gradients at later time points or further away from the source. 

The authors nonetheless insist on the key conclusion from their paper, that morphogen gradients are too noisy to specify cell fate in the NT beyond the first 30\,h. They argue that their analysis must be considered correct because it 1) provides an explanation for the sensitivity of the progenitor markers to both SHH and BMP, and 2) is consistent with their previous postulate that cell differentiation rather than morphogen gradients defines the progenitor domain boundary positions at later stages \cite{Kicheva:2014}. However, one can imagine other roles than precision for this parallel SHH/BMP input \cite{Vetter:2023}, and their own work \cite{Kicheva:2014} and that by others \cite{Yu:2013} showed that the key marker of the motor neuron domain, OLIG2, remains sensitive to SHH signaling also at later stages. The limits posed by 8-bit imaging and the inaccuracy of the chosen error approximation are rigorous mathematical facts. They cannot be challenged biologically. 

Given that the experimental data cannot be used beyond the 8-bit limit, we developed computational approaches to estimate the gradient variability further away from the source based on available experimental measurements. This necessitated assumptions, which the authors now question. For one, based on error propagation, we determined the expected gradient variability beyond the 8-bit detection limit for the case that gradients maintain their measured exponential shape across the domain. Secondly, we developed a cell-based simulation framework that allows us to estimate the positional error from measured molecular noise levels. Both approaches showed that gradients remain sufficiently precise such that gradients can, in principle, pattern the NT throughout its development. 

Zagorski et al.\ claim that it is unrealistic to assume that gradients remain exponential.\ However, they use exponential gradients, rather than their measured gradients, in their own paper \cite{Zagorski:2017}, both in their error propagation method (FitEPM), and also as input when evaluating the potential of opposing gradients in NT patterning via their decoding map, as the reported gradients are too flat and noisy outside the 8-bit limit, also when both gradients are considered simultaneously. As we showed in follow-up work, our results also apply to non-exponential gradients \cite{Adelmann:2023}, and gradient precision is substantially higher still when taking into consideration that morphogens spread at least in 2D, rather than 1D \cite{Long:2013}. 

\begin{figure*}
	\centering
	\includegraphics[width=0.667\textwidth]{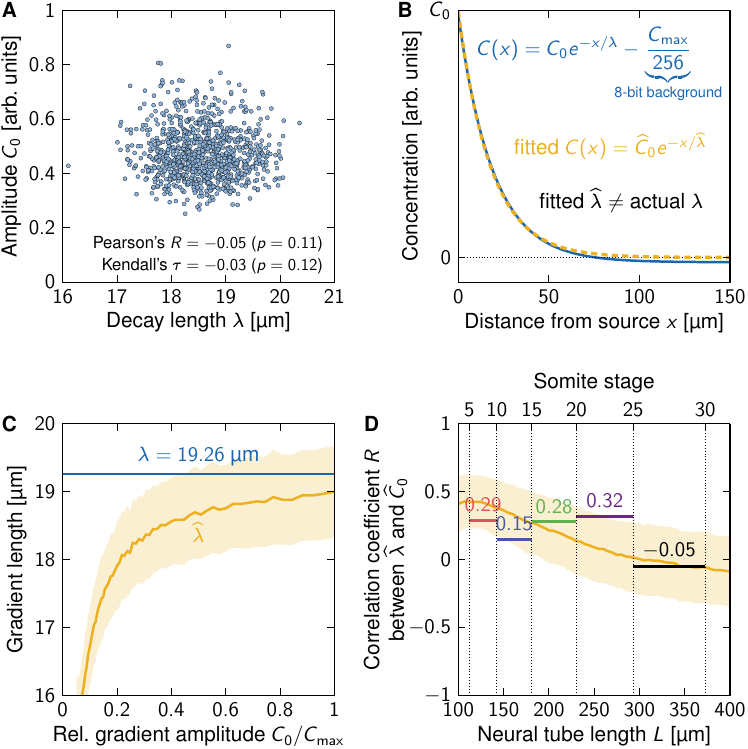}
	\caption{\textbf{Gradient amplitude and decay length are uncorrelated.}
\textbf{A} Absence of correlation between gradient amplitude and decay length in our cell-based model with independent cell-to-cell molecular noise. Data from $n=1000$ simulated gradients obtained with independent noise between individual cells in all three kinetic parameters, $\mathrm{CV}_{p,d,D}=0.3$. See \cite{Vetter:2022} for method details.
\textbf{B} Fitting an exponential (orange) to an exponential gradient from which a background was subtracted (blue, here one 256th of the maximum signal $C_\mathrm{max}$ due to the 8-bit imaging depth used by Zagorski et al.) leads to an error in the apparent gradient decay length that depends on the gradient amplitude. For the shown example, $C_0/C_\mathrm{max}=0.2$.
\textbf{C} We demonstrate this with $n=1000$ synthetic noisy exponential gradients per value of $C_0$, with equal $\lambda=19.26$\,\textmu{}m but different $C_0$ ($C_0=\exp\left[(L/\text{\textmu{}m}-400)/118\right]$ as in Fig.~7D of \cite{Vetter:2022}), with independent Gaussian noise at a resolution of 1\,\textmu{}m. The noise level in the gradients was set to $\mathrm{CV}_C=0.11$ as inferred from the gradients with kinetic noise at $\mathrm{CV}_{p,d,D}=0.3$ from \cite{Vetter:2022}.
The apparent (i.e., fitted) decay length tends to be underestimated (orange, mean\,$\pm$\,SD), and correlates with the gradient amplitude relative to the maximum signal.
\textbf{D} This spurious correlation, as quantified by Pearson's coefficient $R$ from the above dataset within NT length windows of 75\,\textmu{}m (orange, mean\,$\pm$\,SD), disappears at later stages when the gradient amplitude increases. The correlation that Zagorski et al.\ observe at early stages ($R=0.15$--$0.32$, colored bars) lies well within the expected range of this fitting artifact. The correlation coefficient they did not show for somite stages 25--30 is also indicated ($R=-0.05$, black).}
	\label{fig:1}
\end{figure*}

The authors further claim that our statistical approach was flawed because gradient amplitude and decay length would be correlated (Pearson's $R=0.26$, their Fig.~\ref{fig:2}). Based on the same dataset \cite{Cohen:2015}, but including all developmental stages, we concluded otherwise (Pearson's $R=-0.0061$, Kendall's $\tau=0.056$ \cite{Vetter:2022}), and our cell-based model shows that such a correlation, while expected in a deterministic setting, is negligibly small with independent cell-to-cell molecular noise (Fig.~\ref{fig:1}A). The weak correlation that the authors now report early in NT development arises as a technical artifact from the limitations of 8-bit gradient imaging (Fig.~\ref{fig:1}B--D). This highlights the pitfalls in gradient parameter inference from imaging data and suggests that the reported gradient variability is likely strongly overestimated, and that the true gradient length is underestimated (Fig.~\ref{fig:1}C).

\begin{figure}
	\centering
	\includegraphics[width=0.6\linewidth]{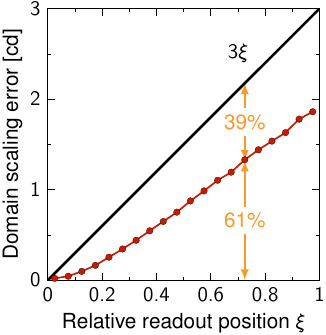}
	\caption{\textbf{Effect of domain rescaling on the positional error estimate in the tissue.}
Based on 50 synthetic gradients, Zagorski et al.\ claim that the scaling error would be $0.15$ cell diameters (cd) at the beginning of NT expansion. However, 50 gradients are not sufficient to obtain precise enough statistics. A reanalysis using $n=\numprint{10000}$ gradients shows that it is $0.32\pm0.01$\,cd (estimate\,$\pm$\,SE) for the example picked by the authors. Evaluating it as described by Zagorski et al., at gradient variability levels as inferred in \cite{Vetter:2022} (Supplementary Information) along the patterning axis (red), shows that the degree of additivity is 61\% (orange). For comparison, the estimate used in \cite{Vetter:2022} is shown (black line).}
	\label{fig:2}
\end{figure}

As we analyzed the processed gradient data that we received from the authors, we noticed that the gradients had been binned from five somite stages and scaled to the same domain length before determining the positional error. This introduces an artificial positional error. We acknowledge that the simple approximation that we used in \cite{Vetter:2022} slightly overestimates the effect as the errors are indeed only partially additive (Fig.~\ref{fig:2}). However, the difference is not particularly important – even uncorrected, the difference between the correctly inferred positional errors of the gradients and those reported for the readouts is rather small and likely reflects, at least in part, remaining technical limitations. We note that the authors overestimate the effect of partial additivity (Fig.~\ref{fig:2}).

Zagorski et al.\ criticize that we did not also correct the readout data. The method section of their Science paper \cite{Zagorski:2017} mentions the scaling neither for the gradients nor for the readout, and the information remained inaccessible to us before publication of our paper. We therefore measured the positional error of the dorsal NKX6.1 boundary ourselves. We could reproduce their results only if we bin, but do not scale the domains, from which we concluded that the authors' readout dataset was likely not scaled. During the writing of our response to their Matters Arising letter, the authors informed us that they did not scale the readout data (A.~Kicheva, personal communication, Supplementary Information). The raw data and scripts that would allow us to check this remain inaccessible to us. While we remain interested in settling the point, we consider domain scaling a minor issue, given the considerable challenges in detecting gradients and aligning them with their readouts, in particular in pseudostratified epithelia, where nuclei are not perfectly aligned with their apical surface \cite{Iber:2022}.\\

In conclusion, none of the points raised by Zagorski et al.\ bear relevance to our conclusions, and remaining uncertainties could be clarified through access to raw data and methodologies. See Supplementary Information for the detailed rebuttal.

\subsection*{Author Contributions}

R.V.\ and D.I.\ jointly reviewed the Matters Arising letter and wrote the reply.

\subsection*{Competing Interests}

The authors declare that they have no competing interests.

\subsection*{Code Availability}

The source code for Figs.~1 and 2 is publicly released under the 3-clause BSD license. It is available as a git repository at \url{https://git.bsse.ethz.ch/iber/Publications/2024_vetter_gradient_variability_ma}.

\clearpage
\renewcommand{\citenumfont}[1]{S#1}
\renewcommand{\bibnumfmt}[1]{[S#1]}
\twocolumn[
\begin{@twocolumnfalse}
\begin{center}
\huge\textbf{Supplementary Information}
\end{center}
\vskip2\baselineskip
\end{@twocolumnfalse}
]{}

In this supplementary document, we respond to the criticism by Zagorski et al.~on our article \cite{SVetter:2022} point by point. The Matters Arising manuscript by the authors is reprinted in bold face, and our response follows each point in Roman.\\

\ccom{\orig{}In the developing neural tube, pattern forms in response to opposing BMP and Shh signaling gradients1. In a recent publication, Vetter and Iber present theoretical analysis based on which they conclude that a single morphogen gradient in the neural tube is sufficient to precisely position gene expression boundaries\textsuperscript{2}. Here we discuss assumptions made by Vetter and Iber that limit the conclusions they reach, and address inaccuracies in their analysis. Given these limitations and existing evidence, it seems likely that both signaling gradients contribute to the precision of pattern formation in the neural tube.}

\reply{}We disagree with the authors that there are inaccuracies in our analysis, as detailed below. On the contrary, our work highlights several inaccuracies in the authors' 2017 Science paper \cite{SZagorski:2017}.\\
 
\ccom{\orig{}In multiple systems, morphogen gradients have been studied by measuring fluorescent reporters of signaling activity in fixed tissues\textsuperscript{3}. A common practice is to estimate the imprecision of a gradient by assessing the variation in fluorescent intensity (FI) between individual embryos at every position in the tissue4. The positional error \boldmath$\sigma_x$ of the gradient is approximated by multiplying the variation of morphogen levels $\sigma_C$ by the local gradient steepness $\abs{\frac{\partial C}{\partial x}}^{-1}$ at that position: $\sigma_x\approx\abs{\frac{\partial C}{\partial x}}^{-1}\sigma_C$. Vetter and Iber point out that different methods for estimating the local gradient steepness can produce different results. One method, numEPM, uses the spatial derivative of mean intensity at the position of interest. Another method, fitEPM, assumes that the mean gradient is exponential. In this case, the local steepness of the gradient is given by the fitted mean intensity at a position divided by fitted exponential decay length. A third method, DEEM, estimates the positional error as the standard deviation of positions $x_{\theta,i}$ that correspond to a defined concentration threshold: $\sigma_x=\SD\{x_{\theta,i}\}$. The DEEM method is derived from the mathematical definition of positional error and hence considered to represent the most direct measure of positional error from an ensemble of gradients.}

\ccom{For low FI values, numEPM and fitEPM methods are influenced by how background FI is estimated and subtracted and by how data is binned and smoothed along the axis. Thus, in the tail of a gradient, the positional error estimates generated by the two methods are inexact and may differ.}

\reply{}Firstly, challenges of background subtraction and smoothening apply also to DEEM. As the authors confirmed (A.~Kicheva, personal communication, Supplementary Information), they employed 8-bit imaging, so that the flat tail of their gradients reflects the detection limit of their chosen imaging depth. Under such circumstances, any computational analysis is meaningless beyond the detection limit.

Secondly, the differences between the computational methods exist independently of any measurement consideration. FitEPM yields the wrong positional error whenever the mean of the gradients deviates from an exponential function. The gradients that have been reported by Zagorski et al.~are fitted well by exponential functions close to the source, and are roughly constant in the center of the domain. In both cases, FitEPM yields the wrong result further away from the source as neither the mean of exponential functions nor the mean of constant functions is an exponential function.\\

\ccom{\orig{}Vetter and Iber claim they can determine which of the two methods is correct by testing which method gives the result closest to estimating the precision of an artificial dataset consisting of an ensemble of exponential gradients using the DEEM method. This leads them to conclude that NumEPM is correct while FitEPM overestimates the positional error. However, this conclusion depends on the assumption that experimental gradients are perfectly exponential.}

\reply{}It is an indispensable pillar of quantitative science to test the accuracy of methods by applying them to problems for which the result is known. If a method fails to provide the correct results for known problems, it is unreasonable to apply it to similar problems, for which the correct answer is not known. In this spirit, we are comparing the three methods (FitEPM, NumEPM, DEEM) using a set of exponential functions, as for those, the result is known. We are not claiming that morphogen gradients are necessarily perfectly exponential. What we are showing is that for exponential functions, FitEPM fails to yield accurate results (except close to the source), and thus it cannot be expected to work for similarly shaped noisy gradients.

To compare the positional errors of morphogen gradients and of their readout, the same method must be used to calculate both of them. The authors chose DEEM to quantify the positional error of the readout. DEEM is unrelated to the shape of the gradients or their readouts. Unlike DEEM, the two other methods, NumEPM and FitEPM, estimate the standard deviation of the position indirectly, via error propagation. We can therefore evaluate whether NumEPM and FitEPM are appropriate by comparing their results to DEEM, when applied to data. As we show, NumEPM gives very similar results as DEEM, while FitEPM greatly overestimates the positional error for sets of exponential gradients that have the same variability in their two parameters (amplitude and gradient length) as reported by Zagorski et al., based on their own fitting of their gradient data with exponential gradients. In fact, the positional error that we obtain with FitEPM for the perfectly exponential gradients is very similar to the results reported by Zagorski et al.~based on the measured, noisy data (Fig.~\ref{fig:1}D).

Further away from the source, the reported gradient shapes lose their exponential shape and become flat. Here, FitEPM performs even worse, as it is explicitly based on the assumption of a continued exponential shape.\\

\ccom{\orig{}The cellular response to the signal and tissue heterogeneities result in gradient shapes that deviate from a perfectly exponential curve\textsuperscript{5,6}.}

\reply{}We agree with this statement, and for non-exponential gradients, the mean gradient will potentially deviate even further from an exponential function, introducing an even greater error when using FitEPM.\\

\ccom{\orig{}The poor signal-to-noise ratio in the gradient tail means that the real shape of gradients in this region cannot be reliably measured.}

\reply{}It is not just that the gradient cannot be reliably measured, but rather not be detected at all, except close to the source. Given the author's use of 8-bit imaging, which limits the detectable signal range to 256-fold, the reported flat gradient ``signal'' is likely just technical noise.

The fact that the measured gradients turn from an exponential shape to a flat noisy shape near the 8-bit limit led us to suspect that the reported poor signal-to-noise ratio is most likely the consequence of an inappropriate imaging regime. The imaging depth is not provided in the methods section of \cite{SZagorski:2017}, and was inaccessible to us prior to the publication of our article. The authors meanwhile confirmed that they had used 8-bit imaging (A.~Kicheva, personal communication, Supplementary Information). It is unclear to us why the authors nonetheless insist that after 30 hours, morphogen gradients are too imprecise to inform patterning in the center of the neural tube — even though the employed 8-bit imaging depth does not allow to image the gradient.\\

\ccom{\orig{}Thus, judging the two methods by comparison to an artificial dataset, which may not represent the true shape of gradients, is misleading. In other words, the performance of a method on an idealized dataset does not determine whether this method will work well on real data which may differ from the idealized dataset.}

\reply{}As we explained above, the same mathematical analysis method needs to be used when evaluating the positional error of gradient and readout. This was not done by Zagorski et al., but a method was used that even with perfectly exponential gradients wrongly yields huge positional errors unless the gradients are identical. Our analysis shows that when the same methods are used for both the gradient and readout, their positional errors are compatible with each other.\\

\ccom{\orig{}More importantly, the analysis of Vetter and Iber indicate that there is in fact very good agreement between the precision estimated by the different methods during the relevant stages of neural tube development (0--15\,ss, corresponding to 0--30\,h). An examination of their Fig 1E shows that the two methods produce identical precision estimates for time points 0--5\,ss. For 10--15\,ss, the estimates are also very similar and diverge only in the gradient tail: DEEM and numEPM estimate 5--6 cell diameters, fitEPM results in 6--8 cell diameters. These positional errors occur at distances \boldmath$>60\%$ tissue length from the morphogen source for GBS-GFP and $>45\%$ for pSmad.}

\reply{}As we pointed out in our article, it lies in the nature of the method that the mean of exponential gradients is fitted well by an exponential function close to the source, but not further away. Our paper does not claim that FitEPM is inaccurate early on or very near the source, but later on and away from the source.

Zagorski et al.~were not able to image the exponential gradients beyond somite stage 15 because they employed 8-bit imaging, which allowed them to detect fluorescence intensity changes only over a 256-fold range. Beyond the 8-bit limit ($\approx110$\,\textmu{}m), their imaging returns technical noise, likely explaining the sharp transition from an exponential to a flat gradient shape. Given this technical detection limit, it is impossible to know whether the reported flat part of the gradient reflects biological reality or the technical limits of 8-bit imaging (although the latter seems more likely to us). No matter which of the two scenarios applies, FitEPM will return a wrong positional error in the flat part of the reported gradients. FitEPM relies on an exponential fit to the mean signal, and this fit is dominated by the higher signal close to the source. As such, FitEPM neither provides a correct estimate of the positional error if the gradients were indeed flat, nor for the case that the exponential gradients continued with the same gradient shape as near the source (but cannot be detected by 8-bit imaging). We analyzed the latter scenario via error propagation methods in our article and showed that the expected positional error of the gradients would be largely consistent with that of the readouts if the gradients indeed remained exponential.

In summary, by employing 8-bit imaging and using FitEPM over the entire spatial and temporal range for the gradients, but DEEM for the readouts, the 2017 Science paper wrongly concludes that single gradients become too imprecise to pattern the center of the neural tube.\\

\ccom{\orig{}This similarity in the estimates at early stages is relevant, because, as we show\textsuperscript{1}, early (before 15\,ss) but not late stage gradients are used to establish pattern.}

\reply{}The authors provide no such evidence in their paper. In fact, in a previous paper \cite{SKicheva:2014}, Kicheva and Briscoe show that late interference with SHH signaling still impacts OLIG2 levels, in agreement with previous work by others \cite{SYu:2013}.\\

\ccom{\orig{}In 1, we derive a decoding map of Shh and BMP signaling using the profiles measured at 5\,ss (Fig.~\ref{fig:2}A and S3A therein). We validate this map with experiments that are independent of how the morphogen signaling gradients were imaged.}

\reply{}The part of our paper that addresses the authors' 2017 Science paper is only concerned with the correct method to compare the positional errors of gradients and their readouts, not with the decoding map. But as the authors bring this up, we emphasize that the decoding map would not work to explain the position of domain boundaries were the authors to use their measured gradients at time points other than the very first. At these later time points, at least one of the two gradient profiles reported by Zagorski et al.~is flat in the part of the domain where the NKX6.1 and PAX3 boundaries are observed. Consequently, their reported gradients could not define the readout position, whether they use a decoding map based on opposing gradients or a direct threshold-based readout of a single gradient. The presented experiments are also consistent with alternative explanations and provide no proof for the authors' decoding map.\\

\ccom{\orig{}We demonstrate that the downstream transcriptional network requires morphogen input for \boldmath$<30$\,h to generate the pattern and this mechanism is sufficient to maintain gene expression in the absence of ongoing signaling at late stages. This reinforces previous experimental evidence, based on growth rate measurements, lineage tracing and perturbation experiments, that indicates that the temporal window for morphogen-dependent cell fate specification is during the first 30\,h of mouse neural tube development\textsuperscript{7}. Thus, for the time interval that is relevant for pattern formation, the fitEPM, NumEPM and DEEM methods produce similar estimates of positional error.}

\reply{}Despite all their quantitative work, the authors have never shown that this transcriptional network yields the correct progenitor domain boundaries when the measured gradients are used as input. Even though Zagorski et al.~claimed above that our use of exponential gradients was unrealistic, they used an idealized exponential gradient, rather than their measured gradients, as input, and did not quantitatively match their boundary data over time. 

As to the claim that morphogen gradients specify patterns only during the first 30\,h of NT development, genetic experiments have shown that removal of the Shh source at later stages affects SHH-dependent NT patterning \cite{SYu:2013}, which should not happen if patterning became independent of morphogen input after 30\,h. The claim that the gradients were too imprecise beyond the first 30\,h hinges on the use of a mathematical flawed method, and on 8-bit imaging that makes it impossible to detect exponential gradients beyond this point, as detailed above.

The conclusion that the rate of differentiation changes rapidly at around 30\,h in the cited publication \cite{SKicheva:2014} hinges on a mathematical analysis that is inaccessible to us and that cannot be reproduced by us with their published data. 

But again, this is of no concern regarding the topic of our analysis, which focuses on a consistent way to compare the positional errors of gradients and their readouts.\\

\ccom{\orig{}Vetter and Iber also argue that imprecision of the signaling gradients is overestimated by grouping signaling profiles into temporal bins that correspond to 10h of developmental time. For a given bin, all signaling profiles are assumed to have the same DV length. Vetter and Iber suggest that this introduces a “scaling error”. To define it, they assume that all profiles in the temporal bin have equal amplitudes and decay lengths, but different absolute lengths. They reason that any resulting positional error is therefore the product of the differing lengths, rather than actual variability in the amplitude and decay length.}

\reply{}This is a misinterpretation of our work. In the quantitative part (Fig.~\ref{fig:2}C--E,H--J), we are using the variability in the amplitude and gradient length that the authors reported in their 2017 Science paper. Only to illustrate the concept of the scaling error in the schematics of Fig.~\ref{fig:2}F,G, the same $C_0$ and $\lambda$ are used for visual clarity. This variability was determined after the gradients had been scaled. We are correcting for this effect by exploiting the reported uniform NT growth behavior. But we agree that this correction can, unfortunately, only be approximate.\\

\ccom{\orig{}This reasoning is problematic. First, if the signaling gradient profiles are corrected in this way, so should the gene expression boundaries of Pax3 and Nkx6.1. Vetter and Iber did not do this. Instead they compare the corrected signaling gradients to the imprecision of Pax3 and Nkx6.1 as reported in Zagorski et al., that is without correction.}

\reply{}The information whether Zagorski et al.~also scaled the NT length in each 5-somite bin before evaluating the PAX3 and NKX6.1 boundaries, and whether they would therefore need to be corrected as well, was inaccessible to us before publication of our paper. When we determined the positional error of the PAX3 and NKX6.1 domain boundaries ourselves using our own imaging data, we could reproduce the reported positional error only when we binned, but did not scale the domains. Having said this, this could, of course, be the result of differences between mouse lines. Based on this finding and the different methods that Zagorski et al.~used to calculate the positional error for gradients and readouts, we concluded that it is most plausible that they measured the PAX3 and NKX6.1 boundaries on unscaled domains.

During the writing of our response to their Matters Arising letter, the authors informed us that they did not scale the readout data (A.~Kicheva, personal communication, Supplementary Information). The raw data and scripts that would allow us to check this remain inaccessible to us.

The remaining difference is in any case small (at most one cell diameter), and we expect that additional effects from the epithelial structure (pseudostratification) amplify the measured positional error \cite{SIber:2022}, and that, when carefully remeasured, the positional errors of both gradients and readouts will turn out to still be lower. As we show in follow-up work, the expected positional error of the morphogen gradients is even lower when considering that they are not spreading on a 1D domain, but at least in 2D \cite{SLong:2013}.\\

\ccom{\orig{}Furthermore, by subtracting the scaling error, Vetter and Iber assume that it has an additive contribution to the overall profile variability. This excludes the possibility that variability in decay length and amplitude could dominate any scaling variability. In such a scenario, subtracting the scaling error would lead to unrealistic underestimation of the actual error (Fig.~\ref{fig:1}). Taken together, the proposed ``scaling error'' correction is applied inconsistently and might underestimate the actual variability.}

\reply{}Given our lack of access to the authors' raw imaging data, the degree of additivity of the scaling error in the authors' methodology on the biological positional error remains unknown to us. However, as we show in our article, the positional error of the gradients and the readout are comparable even if the errors are only partially additive.\\

\ccom{\orig{}Vetter and Iber suggest that gene expression boundaries in the neural tube are positioned by a single morphogen gradient, rather than the combined interpretation of both signaling pathways. }

\reply{}Our work shows that single gradients can provide sufficient positional information in the neural tube. We do not claim that other precision-enhancing effects are excluded, nor that gene expression boundaries are positioned by just one of the morphogen gradients in the neural tube.\\

\ccom{\orig{}Implicit in this idea is that cells somehow distinguish which of two independent gradients is the most precise and use that to determine their identity.}

\reply{}We do not make such an assumption. It appears to us that the authors might have misinterpreted our Fig.~6A,D,E to arrive at this impression. In Fig.~6B,D,E we explicitly show that such an assumption is not necessary.\\

\ccom{\orig{}This interpretation also misses a crucial point: there is experimental evidence that neural progenitors respond to combinations of signaling factors. Consistent with prior studies\textsuperscript{8}, we1 show that neural progenitor identities depend on the levels of both BMP and Shh signaling.}

\reply{}We do not challenge these experimental observations. However, there is no evidence that they serve to increase precision. In fact, as shown by others \cite{SMizutani:2006}, also the ventral-most NKX2.2 domain responds to BMP, even though according to the analysis by Zagorski et al., the BMP gradient would be way too noisy to yield any useful positional information there.\\

\ccom{\orig{}Vetter and Iber further suggest that gradient variability can be accurately inferred from ``summary statistics of exponential gradients''. This necessitates several assumptions. First, gradients are assumed to be exponential. However, diffusion and degradation often depend on feedback from morphogen signaling, which can lead to deviations from exponential shape\textsuperscript{9}.}

\reply{}In this second part of our paper, we developed an error propagation formula to estimate the positional error far from the source (where measurements have so far reached technical limits) based on the variability that can be measured near the source. Whether or not gradients are exponential is indeed an open question. For those cases, in which gradients do remain exponential, our formula can be used. Recognising that they may not necessarily need to do so in other systems of gradient-based patterning, we also presented a simulation framework in the final part of our paper that can be used to estimate the positional error also for more complex cases, as the authors refer to. Using this simulation framework, we have recently shown that non-linear decay, which results in non-exponential gradients, and which had previously been proposed to increase the robustness to variability in the source \cite{SEldar:2003}, yields very similar results as linear decay, and does not result in a relevant increase in precision \cite{SAdelmann:2023}. Morphogen spreading in 2D rather than in 1D, however, substantially reduces gradient variability and therefore permits subcellular precision for physiological levels of molecular noise \cite{SLong:2013}. Together, this reinforces our previous findings that single morphogen gradients can be sufficiently precise to pattern the mouse neural tube, also when considering additional physiological aspects.\\

\ccom{\orig{}Second, ligand and signaling gradients are assumed to have comparable variability and any discrepancy results from technical measurement errors. This ignores the possibility that the signal transduction mechanisms alter the noise properties of a signal\textsuperscript{10}.}

\reply{}We do not make such an assumption. However, if the readout was more precise than the gradients, then the missing information would have to be introduced somehow. Zagorski et al.~proposed that this missing information is obtained by reading out both gradients simultaneously in the center of the domain. As we show, single gradients can be sufficiently precise to define the progenitor domain boundaries also in the center of the NT, offering a simple explanation of how patterning is controlled in the NT.

Zagorski et al.~themselves make the assumption that the GBS-GFP and pSMAD gradients can serve as proxies for the SHH and BMP gradients, even though GBS-GFP may respond very differently from more dorsal SHH-dependent genes, as we explained in detail in our paper.\\

\ccom{\orig{}Third, variables, such as \boldmath$C_0$ and $\lambda$, are assumed to be independent and uncorrelated. Given that both $C_0$ and $\lambda$ depend on the diffusion coefficient and degradation rate, this assumption can easily be violated. Indeed assessing the correlation between $C_0$ and $\lambda$ for measurements taken from 5--25\,ss embryos reveals a modest but significant correlation of $R=0.26$ (Pearson correlation coefficient; $p=0.001$) (Fig.~\ref{fig:2}). This is inconsistent with the assumption that $C_0$ and $\lambda$ vary independently.}

\reply{}We have checked this in our paper using the same dataset \cite{SCohen:2015} but over the entire time course rather than limited to the first 25 somite stages, and concluded otherwise (Fig.~4G in \cite{SVetter:2022}). As the data was plotted using vector graphics, we could extract the plotted point pairs, from which we determined Pearson's $R=-0.0061$ ($p=0.94$) and Kendall's $\tau=0.056$ ($p=0.26$), suggesting that any correlation between $C_0$ and $\lambda$, if it exists, is negligible.

This is confirmed by our cell-based simulations (Fig.~\ref{fig:1}A): The values of $C_0$ and $\lambda$, that we determined from fitting the numerically simulated gradients with independent noise in the three kinetic parameters, are largely uncorrelated: $R=-0.05$ ($p=0.11$) and $\tau=-0.03$ ($p=0.12$). This confirms the validity of our assumption.

The correlation that the authors observe at early stages ($R=0.15$--$0.32$) lies within the expected range of a fitting artifact arising from the technical limitations of 8-bit imaging, as detailed in our main reply (Fig.~\ref{fig:1}B--D). Thus, the author's data provides no evidence for real correlation.\\

\ccom{\orig{}In conclusion, the assumptions inherent to the work of Vetter and Iber and their decision not to take into account experimental evidence make their conclusion, that gene expression boundaries in the neural tube are accurately positioned by a single morphogen gradient, unconvincing.}

\reply{}None of the points of criticism raised by Zagorski et al.~bears relevance to the conclusions of our article, and as pointed out above, we clearly state that our work shows that single gradients can, in principle, be precise enough to encode patterns in the neural tube. We will discuss how the progenitor domains are defined in forthcoming work.

In summary, we demonstrated that the conclusion of Zagorski et al., that the positional error of gradients is much higher than that of their readouts, is the consequence of using different methods to calculate them, and of using 8-bit imaging that can detect fluorescent signals only over a 256-fold range. When using consistent methods, the positional errors are very similar within the distance from the source where 8-bit imaging can yield technically sound results.\\

\subsection*{Figure Legends:}

\noindent\ccom{\orig{}Fig.~\ref{fig:1}. A numerical example of the rescaling error for exponential gradients with variable \boldmath$C_0$ and $\lambda$.}

\noindent\ccom{A. Left: Set of 50 randomly generated exponential morphogen profiles \boldmath$C(x) = C_0e^{-x/\lambda}$. Mean $\lambda=20$\,\textmu{}m and $C_0 = 1$. $\lambda$ and $C_0$ were varied by adding Gaussian noise with $\mathrm{CV}_\lambda=0.2$ and $\mathrm{CV}_{C_0}=0.2$. The domain length was randomly selected from uniform distribution between min $L = 100$\,\textmu{}m and max $L = 150$\,\textmu{}m. 1 cell diameter (cd) $= 4.9$\,\textmu{}m (as in 1). Dashed horizontal line indicates a concentration threshold $C_\theta = 0.1$. At this threshold, the histogram of positions is shown, the mean position $\mean\{x_{i,\theta}\}$ is 48.4 \textmu{}m from the source, and the positional error is $\std\{x_{i,\theta}\} = 9.3\,\text{\textmu{}m} = 1.90$\,cd.}

\noindent\ccom{Right: The profiles are rescaled to the average length \boldmath$\mean\{L_i\} = 125.7$\,\textmu{}m by rescaling each $\lambda_i$ by a factor $\mean\{L_i\}/L_i$. In this set, $\mean\{x_{i,\theta}\} = 48.8$\,\textmu{}m, and $\std\{x_{i,\theta}\} = 10.0\,\text{\textmu{}m} = 2.05$\,cd. This indicates that rescaling changed the positional error estimate $\std\{x_{i,\theta}\}$ by 0.15\,cd.}

\noindent\ccom{B. In Vetter and Iber, a scaling correction is estimated for exponential profiles without variability. A set of 50 such profiles is shown without and with rescaling (left and right, respectively). Assuming a uniform distribution of values at any given concentration threshold, the scaling error increases with distance to the source and reaches a maximum of 3\,cd. Thus, the scaling error corresponds to \boldmath$3\xi$, where $\xi$ denotes the relative position of the bin from the source. For the mean position at $C_\theta$ considered here ($\mean\{x_{i,\theta}\} = 46.7$\,\textmu{}m), $\xi = 46.7/125.7=0.37$, hence the implied scaling correction is 1.11\,cd ($=5.4$\,\textmu{}m). This is much higher than the rescaling error of 0.15\,cd that we obtained for the dataset in A which incorporates realistic variability in $C_0$ and $\lambda$. Note that for an opposing gradient using the same coordinate system, the scaling correction should be $3(1 - \xi)$. Yet, in Fig.~\ref{fig:2}C and E of Vetter and Iber\textsuperscript{2}, the same correction of $3(1 - \xi)$ is incorrectly applied to both the GBS-GFP and pSmad gradients. Had we used $3(1 - \xi)$ as in 2, the implied correction would be 1.8\,cd and be even more overestimated compared to the actual rescaling error.}

\reply{}50 gradients are not sufficient to obtain precise enough statistics. We repeated the authors' analysis with $n=\numprint{10000}$ gradients and obtained a scaling correction of $0.32\pm0.01$\,cd (estimate\,$\pm$\,SE) instead of 0.15\,cd as purported. The scaling error is thus greater than the authors claim. Moreover, the example mentioned by the authors is specific to variability in the gradient parameters that is likely dominated by technical limitations ($\mathrm{CV}_{\lambda,C_0}=0.2$), as we showed in our reply here (Fig.~\ref{fig:1}C) and in \cite{SVetter:2022}. Repeating the same analysis along the entire domain, but taking as gradient variability the relationships inferred for molecular noise at $\mathrm{CV}_{p,d,D}=0.3$ (Fig.~8F,I in \cite{SVetter:2022}),
\begin{equation*}
\mathrm{CV}_\lambda = \sqrt{\frac{0.270\,\text{\textmu{}m}}{L}}
\end{equation*}
and
\begin{equation*}
\mathrm{CV}_{C_0} = 0.1773 + \frac{L}{\numprint{12871}\,\text{\textmu{}m}} + \left(\frac{L}{5659\,\text{\textmu{}m}}\right)^2,
\end{equation*}
yields domain scaling errors whose purely additive contribution is about 61\% of the $3\xi$ approximation we had used (Fig.~\ref{fig:2}). In light of this new data, the scaling correction to the positional error estimated in \cite{SZagorski:2017} is likely somewhat smaller than shown in Fig.~\ref{fig:2} in \cite{SVetter:2022}. However, the corrected positional errors were previously even lower than the reported ones. We had included the analysis with a simple rough estimate for the scaling correction to address the remaining minor gap between the positional errors inferred for the gradients and the one that had been reported for the readouts. Domain scaling is a subordinate issue in comparison to the other limitations in \cite{SZagorski:2017}.\\

\noindent\ccom{\orig{}Fig.~\ref{fig:2}. Correlation between the amplitude \boldmath$C_0$ and decay length $\lambda$ of measured Shh gradients. $C_0$ and $\lambda$ are obtained from exponential fits to the measured Shh ligand gradients from Cohen et al\textsuperscript{11}. Here, Shh profiles were assigned to developmental stages (designated ss for somite stage) based on their DV length as described in Zagorski et al. Dashed lines are linear fits to the data. For each stage, the Pearson correlation coefficient R is shown in the plot. For the pooled set of profiles between ss5 and ss25, the correlation coefficient is 0.26. Only stages up to 25\,ss are shown.}

\reply{}The apparent correlation between gradient amplitude and decay length can be accounted for by technical limitations when the gradient amplitude is still small. For details, see our reply in the main text and Fig.~\ref{fig:1}.

\begingroup
\section*{Supplementary References}
\renewcommand{\section}[2]{}%

\endgroup



\section*{Supplementary material}

## Data & methods neither available to us, nor to the scientific community

The following is a list of missing data and method details that would be needed to clarify remaining open questions regarding our and our colleagues' work.

|   | Data / Information | Reason / Open question | Data source | Response |
|---|---|---|---|---|
| 1 | Individual unprocessed raw gradients of GBS-GFP and pSmad. | To directly determine the positional error for the gradients, and to determine the scaling error introduced by rescaling the gradients to mean lengths in each bin. | [1] | We are sending the raw data per email alongside with this file. |
| 2 | Individual unprocessed raw images from which the gradients in point 1 were measured. | To check the imaging depth used. | [1] | The imaging depth is 8-bit |
| 3 | Source code or complete mathematical description of how averaged gradient decay lengths and amplitudes were calculated for GBS-GFP and pSmad. | We are not able to reproduce the reported averages and standard deviations for lambda and C0 in Fig. 1B,C in [1]. | [1] | We are sending the source code per email alongside with this file. |
| 4 | Information whether or not the patterning domains were scaled in the quantification of positional error of readout boundaries. | To settle whether the domains were scaled only in case of the gradients or also in case of the readouts prior to the determination of the positional error. | [1] | The target gene profiles were treated the same way as GBS-GFP and pSmad profiles, as described in Supplementary materials of [1]: "The fluorescence intensity profiles were then recorded and processed as described (9). The most dorsal and most ventral 5% of L correspond to the approximate positions of the Shh and BMP source boundaries (floor plate and roof plate, respectively) and were therefore excluded from the analysis. The profiles with spatial coordinate x rescaled to L, were used in further analyses." |
| 5 | Mathematical formula for the calculation of the differentiation rate from the measured numbers of progenitors and neurons as reported in Fig. 3C in [2]. | The authors argue that the findings in [2] support the main conclusion in [1], but the formula is not available, and we cannot reproduce the reported differentiation rate unless we use a mathematically flawed approach. | [2] | It is provided in Materials and methods of [2], section "Progenitor and neuron numbers and differentiation rate". |

## Data & methods available to us, but not to the scientific community

The following is a list of data and method details that have been made available to us in private communication, but which are currently not available to the general scientific community.

|   | Data / Information | Reason / Open question | Data source |
|---|---|---|---|
| 1 | Processed gradients of GBS-GFP and pSmad. | We were not allowed to make these data files available to the scientific community, and colleagues can thus not double-check and build on this aspect of our work. | [1] |
| 2 | Source code or complete mathematical description of how the positional error of the GBS-GFP and pSmad gradients as shown in Fig. 1E of [1] was computed from the individual gradients. | While James Briscoe confirmed to us that they used a method corresponding to FitEPM, the method details have not been published by the authors themselves. | [1] |



\end{document}